\documentclass{aa}
\usepackage{graphics, epsfig}
\begin{document}

\title{The lensing properties of the Sersic model}

\author{V. F. Cardone}

\offprints{{\tt winny@na.infn.it}}

\institute{Dipartimento di Fisica ``E.R. Caianiello'', Universit{\`{a}} di Salerno and INFN, Sezione di Napoli, Gruppo Collegato di Salerno, Via S. Allende, 84081 - Baronissi (Salerno), Italy}

\date{Receveid / Accepted }

\abstract{It is well known that the surface brightness of elliptical galaxies and of bulges of spiral galaxies is best fitted by the Sersic $r^{1/n}$ profile. It is thus interesting to explore the lensing properties of the Sersic model because of its wide range of applicability. To this aim, we evaluate the lensing potential, the deflection angle, the time delay between the images and the amplification for a circularly symmetric lens whose surface density is described by the Sersic profile. We estimate the same quantities also adding a shear term to the deflection potential in order to study (at the lowest order) the impact of deviations from radial symmetry or the contribution from other nearby lenses. Moreover, we investigate the systematic errors due to the use of a de\,Vaucouleurs profile instead of the correct Sersic model also taking into account the presence of a dark halo.

}

\titlerunning{The lensing properties of the Sersic model}

\maketitle

\section{Introduction}

Elliptical galaxies present a striking regularity in their global luminosity distributions in the sense that, within a wide range of sizes, their surface brightness profiles can be described by simple functions. The de Vaucouleurs or $r^{1/4}$\,law (\cite{deV48}) was first proposed as a quite general function to fit the light profile of the elliptical galaxies and of the bulges of lenticular and spiral galaxies. However, it was soon realized that the Sersic $r^{1/n}$ profile, which generalizes the de Vaucouleurs law, is best suited to describe the surface brightness distribution of these systems (\cite{CCD93,GC97,PS97}). The regular properties of ellipticals have been the subject of different approaches, concerning both photometric and spectroscopic parameters, resulting in interesting scaling relations. Some well known examples are the fundamental plane relating the effective radius $r_e$, the luminosity intensity $I_e$ at $r_e$ and the central velocity dispersion $\sigma_0$ (\cite{DD87,7S87}); the photometric plane linking the three parameters of the Sersic profile (\cite{KWKM00,G02}) and the entropic plane (\cite{Metal01}). 

The regularity of elliptical galaxies is an attractive feature for cosmological applications of gravitational lensing since it makes it possible to reduce the degeneracy among lens parameters and hence the systematic uncertainty in the estimate of cosmological quantities. The time delay among the images in multiply imaged quasars systems may be used as a tool to determine the Hubble constant $H_0$ (\cite{Ref64}) once a model for the lens has been fitted to the observed images configuration, flux and time delay ratios. The use of scaling relations makes it possible to narrow the parameter space expressing some of them as function of the remaining ones. On the other hand, it has been estimated that at least 80$\%$ of gravitational lenses are elliptical galaxies (\cite{TOG84,FT91}). Hence, it is important that statistical lensing studies, such as the number counts of lens systems, the distribution of image angular separations and of time delays, take into account the most correct model of elliptical galaxies in the analysis of the available data. Up to now, this has not been done since these studies have been concentrated on the dark halos rather than on the luminous components of the galaxies. Therefore, this latter component is usually neglected or simply modeled with the spherical Hernquist profile (\cite{H90}) which approximately reproduces the de\,Vaucouleurs law when projected to the lens plane. The only remarkable exception is the work by Maoz \& Rix (1993) where the lens is modeled using the $r^{1/4}$\,law directly, together with an isothermal dark halo. It should be interesting to repeat their analysis using the more realistic Sersic profile and the scaling relations we have quoted above. As a first step, one has to study the lensing properties of this model. This is the aim of the current work.

The paper is organized as follows. In Sect.\,2, we determine the lensing potential and the deflection angle for a circularly symmetric lens described by the Sersic model. We are thus able to write the lens equations whose solutions are found in Sect.\,3. Section\,4 is devoted to the evaluation of the amplification of the images, while the effect of a shear term, due to deviations from circular symmetry or to nearby lenses, is investigated in Sect.\,5. A study of the systematic errors due to the use of a de\,Vaucouleurs profile mimicking the best\,-\,fit Sersic model to the lens surface brightness is presented in Sect.\,6. We summarize and conclude in Sect.\,7.

\section{Lensing potential and deflection angle}

The surface mass density of a galaxy whose surface brightness is fitted by the Sersic model may be written as (\cite{Sersic})\,:

\begin{equation}
\Sigma(r) = \Upsilon \ I_e \exp{\left \{ - b(n) \left [ \left ( \frac{r}{r_e} \right )^{1/n} - 1 \right ] \right \}}
\label{eq: sigma}
\end{equation}
in which $\Upsilon$ is the mass\,-\,to\,-\,light ratio, $I_e$ the luminosity density at the effective radius $r_e$ and $b(n)$ a constant defined such that the luminosity within $r_e$ is half the total luminosity. It is possible to show that $b(n)$ may be found by solving the following equation (see, e.g., \cite{MC02})\,:

\begin{equation}
\Gamma(2n, b) = \Gamma(2n)/2
\label{eq: bn}
\end{equation}
where $\Gamma(a, z)$ is the incomplete $\Gamma$ function and $\Gamma(a)$ the actual $\Gamma$ function. 

Adopting a rectangular coordinate system in the lens plane centred on the lens galaxy and with $(x_1, x_2)$ oriented along its main axes, $r$ is simply $(x_1^2 + x_2^2)^{1/2}$. We will assume that the surface density is circularly symmetric so that all the lensing quantities will depend only on $r$. While useful in the computations, this approximation is not a serious limitation to our analysis since the results for the circular case may be immediately generalized to flattened models by means of numerical integration (\cite{Schramm,K01}). Moreover, we will investigate later the impact of deviations from circular symmetry by adding a shear term to the lensing potential. However, we stress that the results for the circularly symmetric models allow us to obtain a picture of the main properties of the Sersic model as a lens.

\begin{figure}
\centering
\resizebox{8.5cm}{!}{\includegraphics{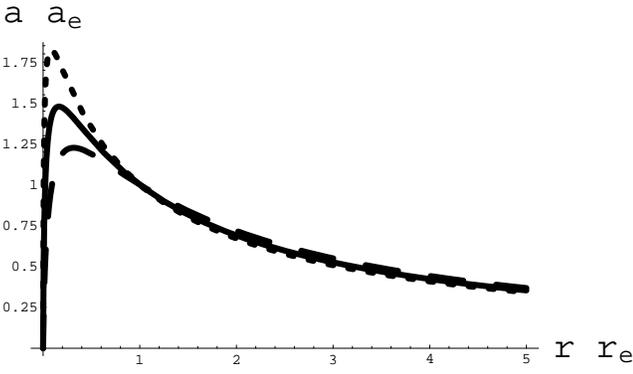}}
\hfill
\caption{The scaled deflection angle $\alpha/\alpha_e$ vs the dimensionless radius $r/r_e$ for three values of the exponent $n$ of the Sersic profile, i.e. $n = 3$ (long dashed), $n = 4$ (solid), $n = 5$ (short dashed).} 
\label{fig: alphavsx}
\end{figure}

It is convenient to introduce a new dimensionless variable defined as\,:

\begin{equation}
x = \left ( \frac{r}{r_e} \right )^{1/n} \ .
\label{eq: defx}
\end{equation}
The lensing potential $\psi(r)$ may be found by solving the following equation (\cite{SEF})\,:

\begin{equation}
\nabla^2 \psi(r) = 2 \kappa(r)
\label{eq: eqpsi}
\end{equation}
where $\kappa(r) = \Sigma/\Sigma_{{\rm crit}}$ is the convergence with $\Sigma_{{\rm crit}} = c^2 D_s/4 \pi G D_l D_{ls}$ the critical density for lensing and $D_s$, $D_l$, $D_{ls}$ are the angular diameter distances between observer and source, observer and lens and lens and source, respectively. Using the above change of variable, Eq.(\ref{eq: eqpsi}) becomes\,:

\begin{equation}
x^{2(1 - n)} \frac{d^2\psi}{dx^2} + x^{1 - 2 n}  \frac{d\psi}{dx} = 
2 n^2 r_e^2 \kappa_e \exp{\left [ - b \left ( x - 1 \right ) \right ]}
\label{eq: eqpsibis}
\end{equation}
in which\,:

\begin{equation}
\kappa_e \equiv  \frac{\Upsilon I_e}{\Sigma_{{\rm crit}}} \ .
\label{eq: defkappae}
\end{equation}
Eq.(\ref{eq: eqpsibis}) may be easily solved using the {\it Mathematica} package. We obtains\,:

\begin{eqnarray}
\psi(x) & = & \kappa_e \frac{{\rm e}^b}{2} x^{2n} {_2F_2[\{2n, 2n\}, \{1 + 2n, 1 + 2n\}, - b x]} \nonumber \\
~ & ~ & - n \ \kappa_e {\rm e}^b \ b^{-2n} r_e^2 \ \Gamma(1 + 2n) \ln{x} + c_1 \ln{x} + c_2 \ ,
\label{eq: psigen}
\end{eqnarray}
where ${_pF_q[\{a_1, \ldots, a_p\}, \{b_1, \ldots, b_q\}, y)}$ is the generalized hypergeometric function\footnote{We use here the same notation for the generalized hypergeometric function as in the {\it Mathematica} package.} (\cite{GR80}) and $c_1$ and $c_2$ are two integration constants that have to be fixed. Since the lensing potential is defined modulo an additive constant with no physical meaning, we can put $c_2 = 0$. As regards $c_1$, we may impose the constraint that the potential is regular both at zero and at infinity, i.e.\,:

\begin{displaymath}
\psi(x) \rightarrow 0 \, \ x \rightarrow \ 0, \ \infty \ .
\end{displaymath}
This condition allows us to choose the constant $c_1$ in such a way that the final expression for the lensing potential turns out to be\,:

\begin{equation}
\psi(x) = \psi_e x^{2n} \ \frac{{_2F_2[\{2n, 2n\}, \{1 + 2n, 1 + 2n\}, -b x]}}
{{_2F_2[\{2n, 2n\}, \{1 + 2n, 1 + 2n\}, -b ]}}
\label{eq: psi}
\end{equation}
with $\psi_e$ the value of the potential for $r = r_e \iff x = 1$, given by\footnote{By using Eq.(\ref{eq: defpsie}), it is possible to eliminate the denominator in Eq.(\ref{eq: psi}). However, we prefer to write the deflection potential as in Eq.(\ref{eq: psi}) so that the exponent $n$ is the only model parameter entering the scaled deflection potential $\psi/\psi_e$.}\,:

\begin{equation}
\psi_e \equiv \frac{{\rm e}^b r_e^2}{2} \kappa_e \ {_2F_2[\{2n, 2n\}, \{1 + 2n, 1 + 2n\}, -b]} \ .
\label{eq: defpsie}
\end{equation}
Given the circular symmetry of the model, the deflection angle will be purely radial and has an amplitude $\alpha$ given by $d\psi/dr$. Using the new variable $x$ and differentiating, we get\,:

\begin{displaymath}
\alpha(x) = \kappa_e r_e b^{-2n} {\rm e}^b x^{-n} \ \left [ \Gamma(1 + 2n) - 2 n \Gamma(2n, bx) \right ]
\end{displaymath}
\begin{displaymath}
\ \ \ \ \ \ \ = 
2 n r_e \kappa_e b^{-2n} {\rm e}^b x^{-n} \ \Gamma(2n) \left [ 1 - \frac{\Gamma(2n, bx)}{\Gamma(2n)} \right ]
\end{displaymath}
having used the relation $\Gamma(1 + 2n) = 2n \Gamma(2n)$. Using Eq.(\ref{eq: bn}), we may finally write the following expression for the deflection angle\,:

\begin{equation}
\alpha(x) = 2 \alpha_e \ x^{-n} \left [ 1 - \frac{\Gamma(2n, b x)}{\Gamma(2n)} \right ]
\label{eq: alpha}
\end{equation}
in which\,:

\begin{equation}
\alpha_e \equiv \alpha(r = r_e) = n r_e \ \kappa_e b^{-2 n} {\rm e}^b \Gamma(2n) \ .
\label{eq: defalphae}
\end{equation}
Note that we are assuming that the distances in the lens plane are measured in $arcsec$ so that $\alpha$ has the dimension of $arcsec$ and the lensing potential of $arcsec^2$. Eq.(\ref{eq: alpha}) should be obtained also directly since, for a circularly symmetric model, it is (\cite{SEF,K01})\,:

\begin{eqnarray}
\alpha(r) & = & \frac{2}{r} \int_0^r{\frac{\Sigma(r')}{\Sigma_{{\rm crit}}} r' dr'} \nonumber \\
~ & = & 2 n r_e \ \kappa_e x^{-2n} \int_{0}^{x}{x'^{2 n - 1} \exp{\left [ -b (x' - 1) \right ]} dx'} \nonumber \\
~ & = &  2 \alpha_e \ x^{-n} \left [ 1 - \frac{\Gamma(2n, b x)}{\Gamma(2n)} \right ] \ .
\label{eq: alphabis}
\end{eqnarray}
which is the same as Eq.(\ref{eq: alpha}). Note that the agreement between the results for $\alpha$ from the two different procedures is an indirect check that our choice of the two integration constants $c_1$ and $c_2$ in Eq.(\ref{eq: psigen}) is indeed correct. It is interesting to observe that only the exponent $n$ of the Sersic profile determines the behaviour with the dimensionless radius $x$ of the scaled deflection angle $\alpha/\alpha_e$, while the effective radius $r_e$, the luminosity density $I_e$ at $r_e$ and the mass\,-\,to\,-\,light ratio $\Upsilon$ enter only as scaling factors. As an example, Fig.\,\ref{fig: alphavsx} shows $\alpha(x)/\alpha_e$ for three values of $n$.

This plot shows two interesting qualitative features. First, we note that $\alpha/\alpha_e$ quickly decreases with $r/r_e$ with the larger values of $n$ going to zero faster. This is an expected result since it simply reflects the fact that the surface mass density of the Sersic model is exponentially decreasing with $r^{1/n}$. On the other hand, we also observe that the shape of $\alpha/\alpha_e$ is almost independent on $n$ for values of $r/r_e \ge 1$ as is apparent from Fig.\,\ref{fig: alphavsx}. We have checked that the plots for larger values of $n$ are more and more similar as $n$ increases, the only difference being the height of the peak.

\begin{figure}
\centering
\resizebox{8.5cm}{!}{\includegraphics{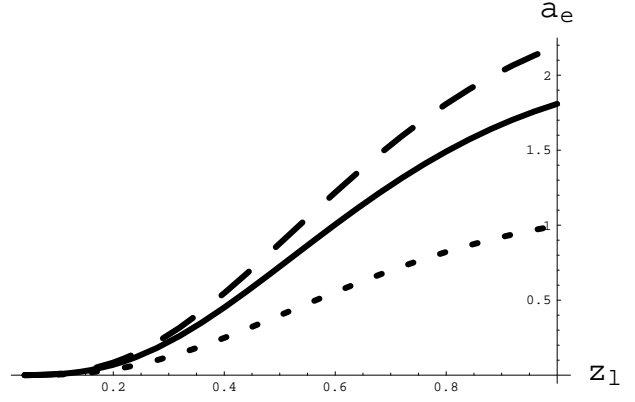}}
\hfill
\caption{The characteristic deflection angle $\alpha_e$ vs the lens redshift $z_l$ for three values of the exponent $n$ of the Sersic profile, i.e. $n = 3$ (long dashed), $n = 4$ (solid), $n = 5$ (short dashed).}
\label{fig: alphaevszl}
\end{figure}

The scaling parameter $\alpha_e$ is a function of the Sersic model parameters $n$, $r_e$ and $I_e$ and of the mass\,-\,to\,-\,light ratio $\Upsilon$ for a given cosmological model and lens $(z_l)$ and source $(z_s)$ redshifts. We observe that $\alpha_e \propto \Upsilon$ as is expected since the higher the lens mass, the higher the bending of light in a given position in the lens plane. To qualitatively study the dependence of $\alpha_e$ on the other model parameters and the lens redshift, it is important to consider lens systems that are as realistic as possible. In Appendix A we give a detailed description of how we choose the Sersic parameters for the lens galaxy taking also into account its redshift. To evaluate $\alpha_e$, we have to compute the critical density $\Sigma_{{\rm crit}}$. To this aim, we need to fix the cosmological parameters and the lens and source redshifts to estimate the angular diameter distances $D_s$, $D_l$, $D_{ls}$. We adopt the flat $\Lambda$CDM model with $(\Omega_m, \Omega_{\Lambda}, h) = (0.3, 0.7, 0.72)$, with $h = H_0/100 \ {\rm km \ s^{-1} \ Mpc^{-1}}$. The source redshift $z_s$ is fixed as $k {\times} z_l$ in which $k \sim 3.7$ is the median value of $z_s/z_l$ among the observed lens systems\footnote{See the CASTLES web page (\cite{CASTLES}).}. Fig.\,\ref{fig: alphaevszl} shows the characteristic deflection angle $\alpha_e$ as function of the lens redshift for three values of $n$ and $\Upsilon = 10$. The increase of $\alpha_e$ with $z_l$ simply reflects the lowering of the critical density because of the higher value of the distances ratio $D_{s}/D_{l} D_{ls}$. Fig.\,\ref{fig: alphaevszl} also shows that the lower $n$ is, the higher is $\alpha_e$. However, it is difficult to give a qualitative explanation for this behaviour. Eq.(\ref{eq: defalphae}) seems to suggest $\alpha_e \propto n b^{-2n} \Gamma(n)$. Actually, the dependence of $\alpha_e$ on $n$ is much more complicated since $n$ also determines $b$ through Eq.(\ref{eq: bn}), while $r_e$ is linked to $n$ and $z_l$ given the way we have chosen the Sersic parameters. These considerations prevent us from giving any (at least) qualitative interpretation of Fig.\,\ref{fig: alphaevszl}.

\section{The lens equation}

\begin{figure}
\centering
\resizebox{8.5cm}{!}{\includegraphics{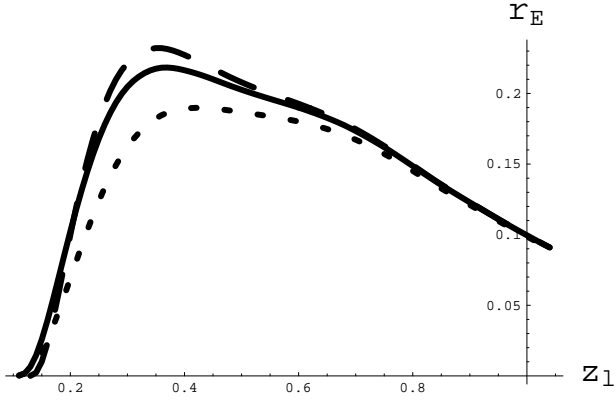}}
\hfill
\caption{The Einstein radius $r_E$ vs the lens redshift $z_l$ for three values of the exponent $n$ of the Sersic profile, i.e. $n = 3$ (long dashed), $n = 4$ (solid), $n = 5$ (short dashed).}
\label{fig: reinstvszl}
\end{figure}

The time delay of a light ray deflected by the galaxy lensing effect is given by\,:

\begin{eqnarray}
\Delta t (r, \theta) & = & h^{-1} \tau_{100} \ {\times} \nonumber \\
~ & ~ & \left [ \frac{1}{2} r^2 - r r_s \cos{(\theta - \theta_s)} + 
\frac{1}{2} r_s^2 - \psi(r, \theta) \right ] 
\label{eq: timedelaygen}
\end{eqnarray}
where $(r, \theta)$ is the image position, $(r_s, \theta_s)$ the unknown source position and $\psi(r, \theta)$ the lensing potential and we have defined\,:

\begin{equation}
\tau_{100} = \left(\frac{D_{l} D_{s}}{D_{ls}}\right) \frac{(1 + z_l)}{c} \ .
\label{eq: taucento}
\end{equation}
According to the Fermat principle, the images lie at the minima of $\Delta t$, so that the lens equations may be simply obtained by minimizing $\Delta t$. Inserting Eq.(\ref{eq: psi}) into Eq.(\ref{eq: timedelaygen}) and differentiating, we get\,:

\begin{equation}
r - r_s \cos{(\theta - \theta_s)} - 2 \alpha_e x^{-n} 
\left [ 1 - \frac{\Gamma(2n, bx)}{\Gamma(2n)} \right ] = 0 \ ,
\label{eq: lenseqa} 
\end{equation}

\begin{equation}
r_s \sin{(\theta - \theta_s)} = 0 \ .
\label{eq: lenseqb}
\end{equation} 
Eq.(\ref{eq: lenseqb}) has two immediate solutions. The first is $r_s = 0$, i.e. the source and the lens are perfectly aligned. In this case, the image is just the Einstein ring with radius $r_E$ obtained solving Eq.(\ref{eq: lenseqa}) for $r_s = 0$. We plot in Fig.\,\ref{fig: reinstvszl} the Einstein radius as a function of the lens redshift for three values of $n$ and $\Upsilon = 10$. Note that the $r_E$ depends significantly on $n$ for lenses with redshift approximately in the range $(0.2, 0.6)$. This is just the redshift range of the most of the observed lens systems (see, e.g., \cite{CASTLES}) thus suggesting that the use of the correct value of $n$ is quite important, i.e. using the de\,Vaucouleurs profile instead of the Sersic model may lead to serious systematic errors. We turn back to this problem in Sect.\,6. 

\begin{figure}
\centering
\resizebox{8.5cm}{!}{\includegraphics{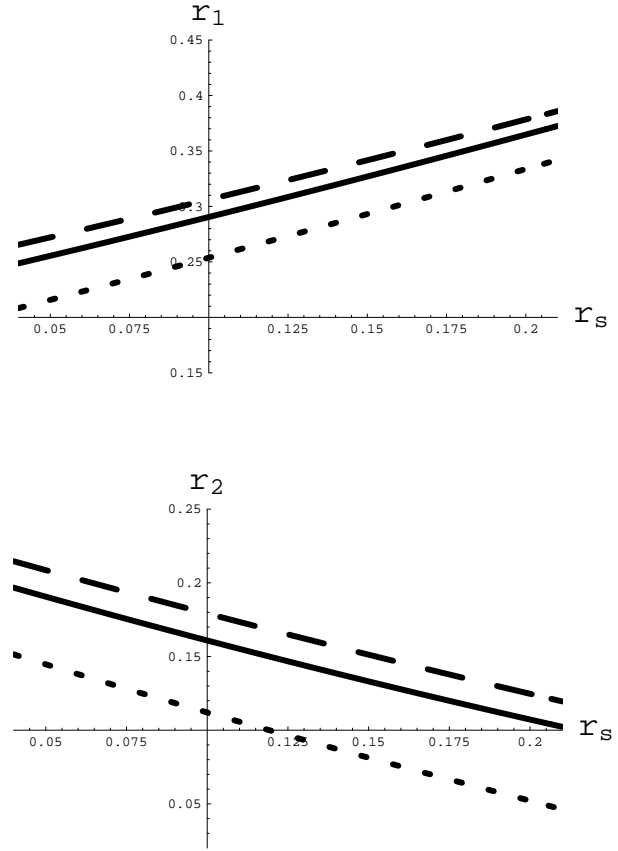}}
\hfill
\caption{The radial coordinates $r_1$ and $r_2$ of the first (top panel) and second (bottom panel) image vs th source position $r_s$ for three values of the exponent $n$ of the Sersic profile, i.e. $n = 3$ (long dashed), $n = 4$ (solid), $n = 5$ (short dashed), with the mass\,-to\,-\,light ratio $\Upsilon = 10$.} 
\label{fig: r1r2vsrs}
\end{figure}

The second solution is obtained by imposing $\sin{(\theta - \theta_s)} = 0$, i.e. $\theta = \theta_s + m \pi$ (with $m = 0, 1$). In this case, we get two images symmetrically disposed with respect to the lens centre. The radial coordinate $r_1$ of the first image (i.e., the one with $\theta = \theta_s$) is obtained by solving Eq.(\ref{eq: lenseqa}) with $\cos{(\theta - \theta_s)} = 1$, while the second has a distance $r_2$ from the lens centre obtained by solving Eq.(\ref{eq: lenseqa}) with $\cos{(\theta - \theta_s)} = -1$. In Fig.\,\ref{fig: r1r2vsrs}, we plot $r_1$ and $r_2$ as function of the source radial position for the different values of $n$ used. The Sersic parameters and the cosmological background are fixed as usual, while we have adopted $(z_l, z_s) = (0.31, 1.72)$, as for the real lens PG\,1115+080 (\cite{W80,Ietal96}).

Note that we get two images which is an even number, while one could invoke the theorem on the odd number of images (\cite{SEF,Straumann}) to suspect that (at least) a third image should appear. However, the theorem is not violated since it only holds if the mass distribution and its first derivative have no singularities. This is not the case for the Sersic model. While $\Sigma(r)$ in Eq.(\ref{eq: sigma}) is regular, its first derivative is singular in the origin and hence the above theorem does not apply.

\begin{figure}
\centering
\resizebox{8.5cm}{!}{\includegraphics{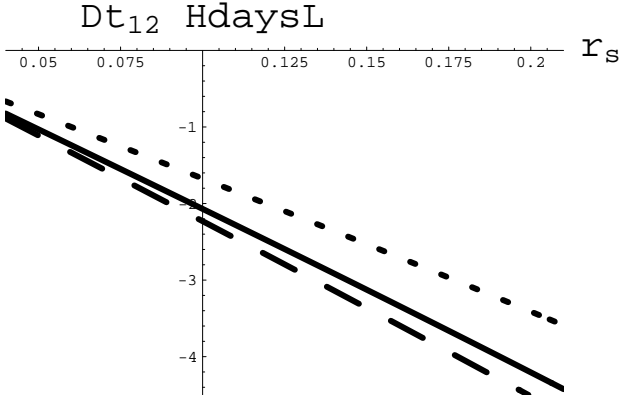}}
\hfill
\caption{Time delay between the two images of a distant source as a function of the source position $r_s$ for three values of the exponent $n$ of the Sersic profile, i.e. $n = 3$ (long dashed), $n = 4$ (solid), $n = 5$ (short dashed).} 
\label{fig: td}
\end{figure}

An important observable in lens systems is the time delay between the two images of a distant source. This may be simply evaluated as\,:

\begin{equation}
\Delta t_{12} = \Delta t(r_1, \theta_1) - \Delta t(r_2, \theta_2) \ .
\label{eq: td12}
\end{equation}
Using Eq.(\ref{eq: timedelaygen}), we may quickly compute the time delay between the two images formed by a circularly symmetric Sersic lens. This is plotted in Fig.\,\ref{fig: td} as function of the source radial coordinate $r_s$ for a system with $(z_l, z_s) = (0.31, 1.72)$ and $\Upsilon = 10$. Note that the time delay scales with the mass\,-\,to\,-\,light ratio as $\Upsilon^2$ since it is proportional to the square of the image positions and the latter (as the Einstein radius) scales linearly with $\Upsilon$. Fig.\,\ref{fig: td} shows that the first image always arrives first being less delayed by the bending of light due to the lens. On the other hand, the time delay is quite short when compared to the values usually measured, the lowest being $\sim 10$\,d. However, drawing any conclusion by this comparison is not possible since one should also take care of the observational problems related to the measure of short time delays.

\section{The amplification of the images}

\begin{figure}
\centering
\resizebox{8.5cm}{!}{\includegraphics{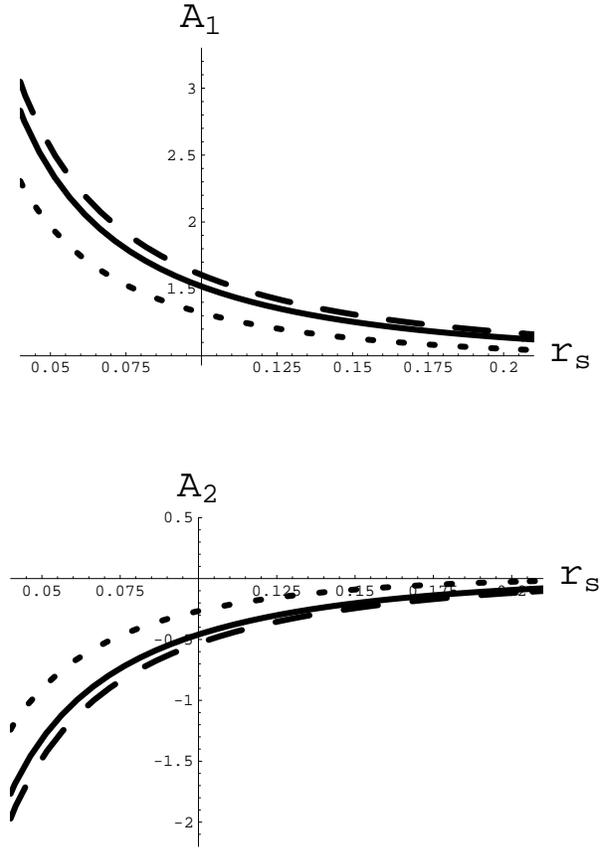}}
\hfill
\caption{The amplification of the first (top panel) and second (bottom panel) image as a function of the source distance from the lens centre for $n = 3$ (long dashed), $n = 4$ (solid), $n = 5$ (short dashed).}
\label{fig: a1a2vsrs}
\end{figure}

The bending of light due to gravitational lensing changes the solid angle subtended by a source on the sky and leads to an amplification of the images with respect to the undeflected case. The image amplification is simply the inverse of the Jacobian of the lens mapping. It is\,:

\begin{equation}
A(x_1, x_2) = \frac{1}{{\rm det} J} = \left [ (1 - \psi_{11}) (1 - \psi_{22}) - \psi_{12}^{2} \right ]^{-1}
\label{eq: ampl}
\end{equation}
in which $\psi_{ij} = \partial^2 \psi/\partial x_i \partial x_j$. For a circularly symmetric lens, this reduces to (\cite{SEF})\,:

\begin{eqnarray}
A(r) & = & \left [ \left ( 1 - \frac{\alpha}{r} \right ) \left ( 1 - \frac{d\alpha}{dr} \right ) \right ]^{-1} \nonumber \\
~ & = & \left [ \left ( 1 - \frac{\alpha}{r_e x^n} \right ) \left ( 1 - \frac{1}{n r_e} x^{1 - n} \frac{d\alpha}{dx} \right ) \right ]^{-1} \ . 
\label{eq: amplrad}
\end{eqnarray}
Inserting Eq.(\ref{eq: alpha}) into Eq.(\ref{eq: amplrad}), with some simple algebra, we finally get\,:

\begin{equation}
A(x) = \left [ (1 - p) (1 + p + q) \right ]^{-1}
\label{eq: amplradbis}
\end{equation}
having defined\,:

\begin{equation}
p(x) \equiv \frac{2 \alpha_e}{r_e} \ x^{-2 n} \ \left [ 1 - \frac{\Gamma(2 n, b x)}{\Gamma(2 n)} \right ] \ ,
\label{eq: defp}
\end{equation}

\begin{equation}
q(x) \equiv \frac{2 \alpha_e}{n r_e} \frac{b^{2 n}}{\Gamma(2 n)} \exp{(-b x)} \ .
\label{eq: defq}
\end{equation}
The amplification of the two images is shown in Fig.\,\ref{fig: a1a2vsrs} where the Sersic model parameters, the cosmological background and the lens and source redshifts have been fixed as for Fig.\,\ref{fig: r1r2vsrs}.

This plot shows that, while the first image is always amplified and conserves the same orientation as the source (i.e., it has positive parity), the second image always has negative parity and could also be deamplified (i.e., the magnification is lower than 1). Therefore, we conclude that for certain values of $n$ and $r_s$ a circular Sersic model is unable to produce two observable images and hence its lensing effect is actually null.

It is interesting to study the critical curves of this model. These are defined as the loci of the points in the lens plane such that the amplification becomes (formally) infinite, while their projection on the source plane gives the caustics. Eq.(\ref{eq: amplradbis}) shows that $A$ diverges either if $1 - p = 0$ or if $1 + p + q = 0$. The first condition individuates the tangential critical curves which, the model being circularly symmetric, will be a circle with radius given by the solution of the equation $1 - p = 0$. Inserting Eq.(\ref{eq: defp}) into this condition, it is easy to show that the only tangential critical curve is just the Einstein ring. The corresponding caustic is thus the point $r_s = 0$. On the other hand, the condition $1 + p + q = 0$ determines the radius of the radial critical curve. However, using Eqs.(\ref{eq: defp}) and (\ref{eq: defq}), this relation reduces to an equation which has no solutions. Thus, we conclude that the circular Sersic model has no radial critical curve.

\section{The effect of the shear}

We have considered the case of a circularly symmetric model, but we are aware that real elliptical galaxies are often flattened systems\footnote{Note, however, that bulges of spiral galaxies are typically more rounder than elliptical galaxies and hence the circular case we have studied is not completely unrealistic.}. If the deviations from circular symmetry are not too large (i.e., if the eccentricity $e = 1 - b/a$ is not too different from 1), the effect of flattening may be taken into account (to the lowest order) adding a shear term to the lensing potential which is now\,:

\begin{equation}
\psi = \psi_{{\rm lens}}(r) - \frac{1}{2} \gamma r^2 \cos{(2 \theta - 2 \theta_{\gamma})}
\label{eq: psishear}
\end{equation}
in which $\psi_{{\rm lens}}(r)$ is given by Eq.(\ref{eq: psi}), while $\gamma$ is the shear intensity and $\theta_{\gamma}$ its orientation angle. A shear term could also mimic the contribution to the lensing potential of nearby lenses or of the cluster (if any) to which the main lens galaxy belongs. Adding the shear term to the lensing potential obviously changes the lens equations. Some simple algebra allows us to finally obtain\,:

\begin{equation}
r = \frac{r_s}{\gamma} \frac{\sin{(\theta - \theta_s)}}{\sin{(2 \theta - 2 \theta_{\gamma})}} \ , 
\label{eq: sheareqa}
\end{equation}

\begin{equation}
[ {\cal{P}}(\theta) - {\cal{Q}}(\theta) ] r_s^2 + {\cal{R}}(\theta) = 0
\label{eq: sheareqb}
\end{equation}
having defined\,:

\begin{figure}
\centering
\resizebox{8.5cm}{!}{\includegraphics{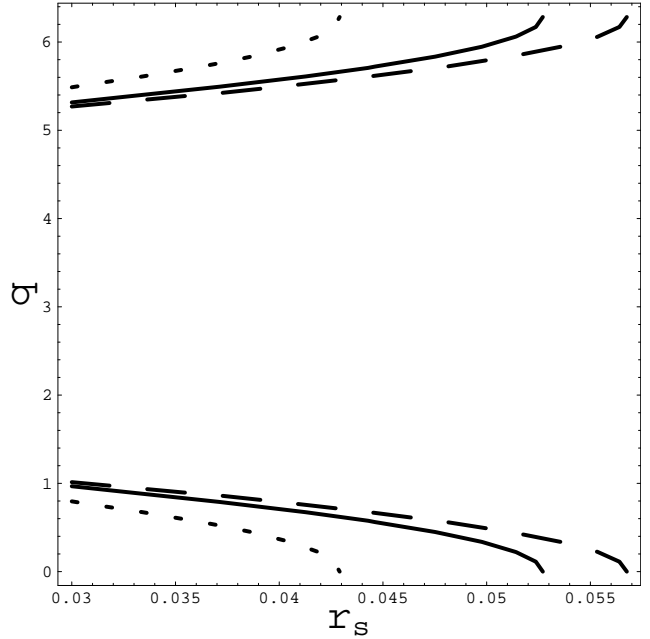}}
\hfill
\caption{Zero\,-\,level contour plot of the left\,-\,hand side of Eq.(\ref{eq: sheareqb}) for a source on axis, i.e. $\theta_s = 0$, as a function of the radial source coordinate $r_s$ (in $arcsec$) and the image angular coordinate $\theta$ (in $rad$) excluding the region giving rise to negative $r$. Results are shown for models with $n = 3$ (long dashed), $n = 4$ (solid), $n = 5$ (short dashed).}
\label{fig: solshearonaxis}
\end{figure}

\begin{equation}
{\cal{P}}(\theta) \equiv \left [ 1 + \gamma \cos{(2 \theta - 2 \theta_{\gamma})} \right ] \ 
\sin^2{(\theta - \theta_s)} \ ,
\label{eq: defpcal}
\end{equation}

\begin{equation}
{\cal{Q}}(\theta) \equiv \gamma \cos{(\theta - \theta_s)} \sin{(\theta - \theta_s)} 
\sin{(2 \theta - 2 \theta_{\gamma})} \ ,
\label{eq: defqcal}
\end{equation}

\begin{equation}
{\cal{R}}(\theta) \equiv 2 \alpha_e r_e \gamma^2 \sin^2{(2 \theta - 2 \theta_{\gamma})} \ \left [
1 - \frac{\Gamma(2n, b x)}{\Gamma(2n)} \right ] \ .
\label{eq: defrcal}
\end{equation}
To find the images position, we can thus solve Eq.(\ref{eq: sheareqb}) with respect to $\theta$ for given values of the source coordinates $(r_s, \theta_s)$ and of the shear parameters $(\gamma, \theta_{\gamma})$ and then use Eq.(\ref{eq: sheareqa}) to find the corresponding value of the radial coordinate, excluding those solutions of Eq.(\ref{eq: sheareqb}) which give rise to negative $r$. Without loss of generality, hereinafter, we will assume $\theta_{\gamma} = 0$. Moreover, we fix $\gamma = 0.11$ as determined by Holder \& Schechter (2003) through a set of numerical simulations\footnote{The value $\gamma = 0.11$ refers to real quadruply imaged systems (\cite{HS02}), while here we do not know in advance what is the number of images. The systematic error we are making using this value does not affect our main results.}. The Sersic model parameters, the background cosmological model and the redshifts of lens and source are fixed as in Sects.\,3 and\,4. Because of the breaking of radial symmetry due to the shear term, the solutions of the lens equations now depend not only on the radial, but also on the angular coordinate of the source. In Fig.\,\ref{fig: solshearonaxis} we show the contour plot at the zero level of the left\,-\,hand side of Eq.(\ref{eq: sheareqb}) as a function of $r_s$ and $\theta$ assuming the source is on\,-\,axis, i.e. $\theta_s = 0$, and excluding the regions which give rise to negative $r$. For a given $r_s$, the solution of Eq.(\ref{eq: sheareqb}) may be graphically found as the intersection points of the vertical line $r_s = const$ with the plotted curves. Two images are formed reducing to only one (the Einstein ring) if $r_s = 0$ (not shown in the plot). When the source is off axis, the number of images is two or four according to the source position, as may be argued looking at Fig.\,\ref{fig: solshearoffaxis} which refers to the case $\theta_s = \pi/3$. 

\begin{figure}
\centering
\resizebox{8.5cm}{!}{\includegraphics{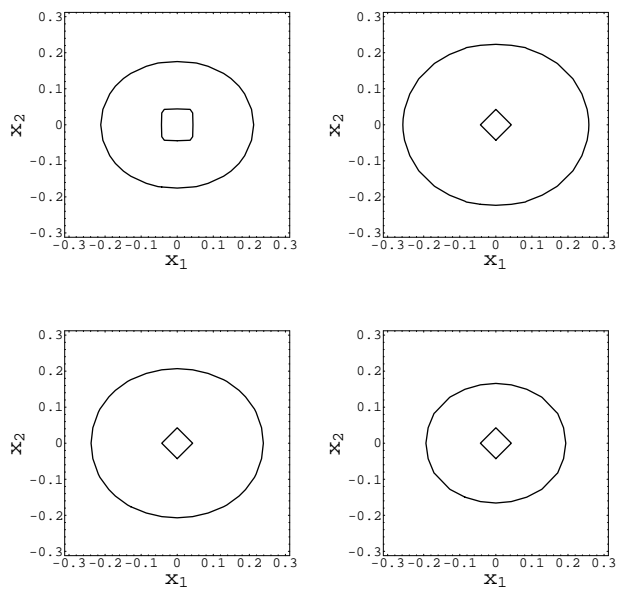}}
\hfill
\caption{Critical curves in the lens plane $(x_1, x_2)$ for the Sersic model modified by the addition of a shear term. The plots refer to $n = 2$ (top left), $3$ (top right), $4$ (bottom left), $5$ (bottom right). Distances in the lens plane are measured in $arcsec$.}
\label{fig: critcurveshear}
\end{figure}

The presence of shear also changes the amplification as is easy to understand observing that, because of the broken circular symmetry, $A$ will now be function of both the polar coordinates $(r, \theta)$. Actually, Eq.(\ref{eq: amplrad}) does not hold anymore and we have to use Eq.(\ref{eq: ampl}) to estimate $A(x_1, x_2)$. As a consequence, the critical curves are changed too. Fig.\,\ref{fig: critcurveshear} shows that now we get two critical curves. The external curve is the tangential critical curve which is now deformed into an ellipse instead of a circle as in the case without shear. Moreover, a second critical curve\footnote{Note that the inner critical curve in the upper left plot in Fig.\,\ref{fig: critcurveshear} is rotated with respect to the other plots. It is important to stress that this is only an artifact of the numerical solution.} appears inside the first corresponding to the radial curve which was absent in the circularly symmetric case. 

\begin{figure}
\centering
\resizebox{8.5cm}{!}{\includegraphics{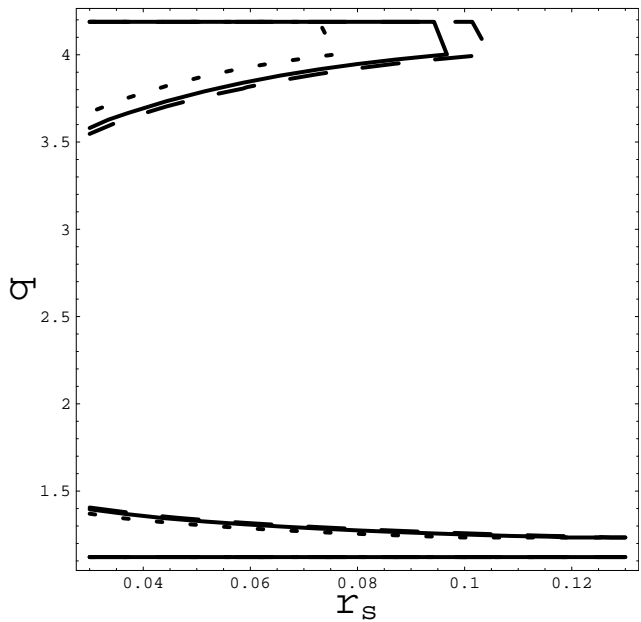}}
\hfill
\caption{Same as Fig.\,\ref{fig: solshearonaxis} but for $\theta_s = \pi/3$. Note that, for graphics\,-\,related problems, the sharp edges in the plot are not real the actual ones being more smoothed. Note that the contours (in particular the lower ones) for different values of $n$ are almost superimposed.}
\label{fig: solshearoffaxis}
\end{figure}

\section{Systematic errors}

Even if the Sersic profile is well known to be a better fit to the surface brightness of elliptical galaxies (\cite{CCD93,GC97,PS97}), it is customary in lensing studies to use the de\,Vaucouleurs $r^{1/4}$ law to model the visible component of the lens. It is thus important to investigate the systematic errors introduced by this procedure. To this end, we proceed as follows. First, we fix the Sersic parameters as described in the Appendix arbitrarily fixing the lens redshift $z_l = 0.31$. As a second step, we generate a simulated set of surface brightness measurements using realistic error bars. These data are then fitted with the de\,Vaucouleurs profile to obtain the values of $r_e$ and $I_e$ that best reproduce the observed surface brightness. This procedure ensures that the de\,Vacouleurs parameters are reasonably chosen and the projected mass distribution of the lens galaxy is well represented both by the true Sersic profile and the fitted $r^{1/4}$ law.

\begin{figure}
\centering
\resizebox{8.5cm}{!}{\includegraphics{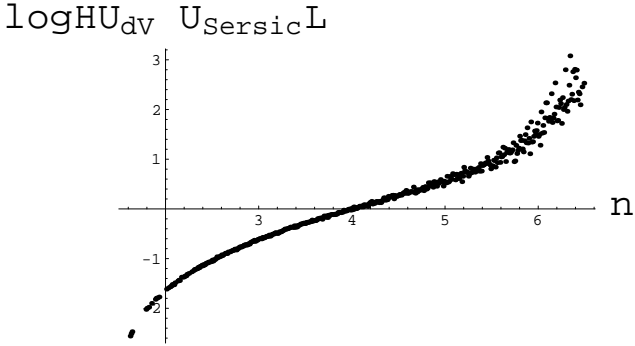}}
\hfill
\caption{The systematic error in the estimate of the mass\,-\,to\,-\,light ratio $\Upsilon$ as a function of the exponent $n$ of the correct Sersic model describing the lens.}
\label{fig: mlerrsers}
\end{figure}

As a first application, we consider the estimate of the mass\,-\,to\,-\,light ratio. It is quite easy to show that\,:

\begin{equation}
\Upsilon = \frac{r_E^{obs}}{r_E(n, r_e, I_e, \Upsilon = 10)} 
\label{eq: estml}
\end{equation}
with $r_E^{obs}$ the observed value of the Einstein ring\footnote{Actually, it is not so easy to observe an Einstein ring, but we may qualitatively estimate $r_E$ from the image separation in multiply imaged quasars.}. Note that we have fixed $\Upsilon = 10$ in the denominator in Eq.(\ref{eq: estml}) just to make the numerical computation easier. Let us suppose now that we have seen an Einstein ring and we have been able to measure the lens surface brightness. Then, we can fit this profile using both the Sersic and de\,Vaucouleurs law so that we know the three parameters $(n, r_e, I_e)$ and may evaluate $\Upsilon$ using Eq.(\ref{eq: estml}). Using the de\,Vaucouleurs law with $n = 4$ instead of the correct Sersic profile leads to a different estimate of $\Upsilon$. Fig.\,\ref{fig: mlerrsers} shows $\log{(\Upsilon_{{\rm dV}}/\Upsilon_{{\rm Sersic}})}$ as a function of the exponent $n$ of the Sersic model used to generate the simulated data, being $\Upsilon_{{\rm dV}}$ ($\Upsilon_{{\rm Sersic}}$) the value obtained modeling the lens with the de\,Vaucouleurs (Sersic) law. We only include in the plot those models that are well fitted by both profiles. 

The results are quite interesting showing that the mass\,-\,to\,-\,light ratio is underestimated if $n < 4$ or overestimated if $n > 4$ by a factor which may be as high as 10 (excluding values of $n < 2$ or $n > 5$ that are less realistic). However, Fig.\,\ref{fig: mlerrsers} should not be considered as a definitive proof against the de\,Vaucouleurs model instead of the Sersic one. Actually, this result have been obtained using simulated observations of lens galaxies whose parameters have been chosen with a procedure that could also be revised. It should be interesting to repeat this analysis for real lens systems also comparing the estimated mass\,-\,to\,-\,light ratios with those predicted by models of stellar evolution.

In the previous discussion, we have assumed that the lens redshift is known which is not always the case. Fig.\,\ref{fig: reinstvszl} shows that the way $r_E$ scales with $z_l$ depends on $n$ too. Henceforth, we should correctly write\,:

\begin{displaymath}
\Upsilon_{{\rm dV}} = \frac{r_E^{{\rm obs}}}{r_E^{{\rm dV}}(n = 4, r_e, I_e, \Upsilon = 10)} \ ,
\end{displaymath}

\begin{displaymath}
\Upsilon_{{\rm Sersic}} = \frac{r_E^{{\rm obs}}}{r_E^{{\rm Sersic}}(n, r_e, I_e, \Upsilon = 10)} \ ,
\end{displaymath}
with $r_E^{{\rm dV}}$ ($r_E^{{\rm Sersic}}$) the Einstein radius evaluated for the de\,Vaucouleurs (Sersic) model. Combining the previous two relations, we get\,:

\begin{equation}
\frac{\Upsilon_{{\rm dV}}}{\Upsilon_{{\rm Sersic}}} = \frac{r_E^{{\rm Sersic}}(n, r_e, I_e, \Upsilon = 10)}{r_E^{{\rm dV}}(n = 4, r_e, I_e, \Upsilon = 10)} \ .
\label{eq: mlratio}
\end{equation}
This depends only on $n$ if we know $z_l$ (as assumed above), otherwise it is a function of both $z_l$ and $n$. This suggests that the results in Fig.\,\ref{fig: mlerrsers} should be read with some caution. Let us consider, for instance, the case $n = 3$. From Fig.\,\ref{fig: mlerrsers} we get $\Upsilon_{{\rm dV}}/\Upsilon_{{\rm Sersic}} < 1$. On the other hand, Fig.\,\ref{fig: reinstvszl} also shows that $r_E^{{\rm Sersic}} < r_E^{{\rm dV}}$ for values of the lens redshift lower than about 0.2. As a result, we conclude that, if $z_l$ were not known, we should be unable to say whether the systematic error is due to having chosen an incorrect lens model or a wrong redshift. Actually, the situation is still more involved since the way we have fixed the Sersic parameters also implictely depends on $z_l$. We thus stress the need to repeat this analysis using real galaxies to thoroughly investigate the systematics induced by using the de\,Vaucouleurs profile instead of the Sersic one. This is outside the scope of this paper, but will be the subject of future work.

Real lensing galaxies are usually modeled adding a dark halo to the visible component. If the Einstein radius has been measured somehow (e.g, by directly observing an Einstein ring or from the images separation), then it is possible to get an estimate of the projected mass of the dark halo inside the Einstein ring as\,:

\begin{equation}
M(r_E) = \pi r_E^2 \ \Sigma_{crit} - M_{{\rm vis}}(r_E) 
\label{eq: massest}
\end{equation} 
with $M_{{\rm vis}}(r_E)$ the projected mass of the visible component estimated from the Sersic (or de\,Vaucouleurs) model as\,:

\begin{equation}
M_{{\rm vis}}(r_E) = 2 \pi \Upsilon n r_e^2 I_e \int_{0}^{x_E}{x^{2n - 1} {\rm e}^{-b(x - 1)} dx} 
\label{eq: massvis}
\end{equation}
with $x_E = (r_E/r_e)^{1/n}$. We may suppose that $(n, r_e, I_e)$ have been measured fitting the Sersic law to the lens surface brightness profile and that $\Upsilon$ has been estimated from a stellar evolution synthesis code so that Eq.(\ref{eq: massest}) may be used to directly obtain the projected mass of the dark halo. 

To investigate the systematic errors induced by the use of a de\,Vaucouleurs profile instead of the correct Sersic one, we first have to choose a dark halo model. We adopt the softened power law sphere (see \cite{K01} and references therein) whose projected surface density is\,:

\begin{equation}
\Sigma(r) = \frac{b^{ 2 -\beta} \ \Sigma_{{\rm crit}}}{2 \ \left( s^2+ r^2 \right)^{1 - \beta/2}} = 
\Sigma_0 \ \left ( 1 + \frac{s^2}{r^2} \right )^{\beta/2 - 1}    
\label{eq: sofkappa}
\end{equation}
which represents a flat core with scale radius $s$ and central surface density $\Sigma_0$, and then a power law decline with exponent $\beta$ defined such that the mass grows as $r^\beta$ asymptotically. The core radius can be zero if $\beta > 0$. As illustrative cases, we consider the two models with $\beta = 1$ (i.e., a softened isothermal model) and $\beta = -2$ (i.e., a Plummer model). To fix the halo parameters, we first deproject both the halo and the Sersic profile (\cite{BT87,MC02}), then estimate the total mass of the visible component and fix the halo virial mass, $M_{{\rm vir}}$, so that $90\%$ of the total galaxy mass is represented by the halo itself. The value of the core radius $s$ is then obtained by scaling that of the Milky Way as $s = s_{{\rm MW}} \ (M_{{\rm vir}}/M_{{\rm vir}}^{{\rm MW}})^{1/3}$. It is then straightforward to evaluate $\Sigma_0$ as function of $M_{{\rm vir}}$ and $s$. 

\begin{figure}
\centering
\resizebox{8.5cm}{!}{\includegraphics{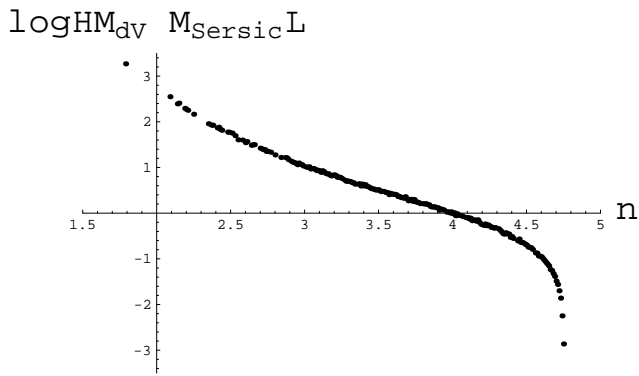}}
\hfill
\caption{The systematic error in the estimate of the projected dark halo mass inside the Einstein radius $r_E$ as a function of the exponent $n$ of the correct Sersic profile using the softened power law model with $\beta = 1$ for the halo.}
\label{fig: masserriso}
\end{figure}

\begin{figure}
\centering
\resizebox{8.5cm}{!}{\includegraphics{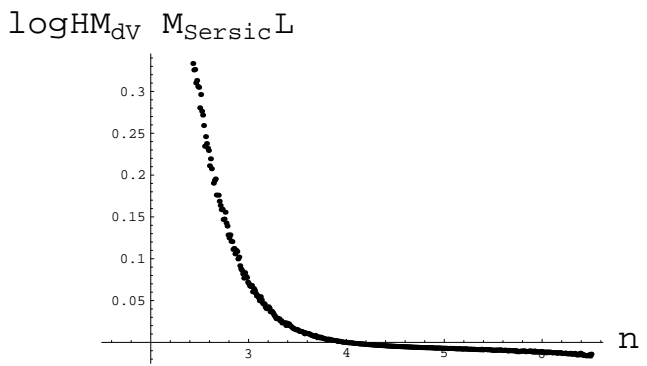}}
\hfill
\caption{Same as Fig.\,\ref{fig: masserriso}, but with $\beta = -2$ for the halo model.}
\label{fig: masserrplum}
\end{figure}

We use Eq.(\ref{eq: massest}) to estimate the projected mass of the dark halo using for the lens visible component the Sersic model or the corresponding fitted de\,Vaucouleurs model assuming $\Upsilon = 10$ for both profiles. The Einstein ring is obtained by solving Eq.(\ref{eq: lenseqa}) with $r_s = 0$ and adding the halo contribution to the lensing potential. In Figs.\,\ref{fig: masserriso} and \ref{fig: masserrplum}, we plot $\log{(M_{{\rm dV}}/M_{{\rm Sersic}})}$ as a function of the exponent $n$ of the correct Sersic model, in which $M_{{\rm dV}}$ ($M_{{\rm Sersic}}$) is the value of the projected mass of the dark halo obtained after the visible component have been fit by the de\,Vaucouleurs (Sersic) profile. As is apparent from Fig.\,\ref{fig: masserriso}, the halo projected mass for the case $\beta = 1$ may be seriously overestimated (underestimated) if $n < 4$ ($n > 4$). Considering only models with $n$ in the range $(3, 5)$ which is more realistic, the error could be up to a factor $10$ so that it is worthwhile to investigate more carefully this topic using real lens systems. 

On the other hand, the error is lower if the halo is modeled as a Plummer sphere as is shown in Fig.\,\ref{fig: masserrplum}. However, the Plummer model is less suited to describe galaxy haloes so that it is not worthwhile to spend further work on this topic.

\section{Conclusions}

Simple analytical calculations show that elliptical galaxies form 80$\%$ of the gravitational lenses. Hence, it is very interesting to investigate their lensing properties taking into account models which are as realistic as possible. The Sersic $r^{1/n}$ profile is well known to fit the surface brightness of both elliptical galaxies and bulges of lenticular and spiral galaxies. In this paper we have studied its lensing properties assuming circular symmetry of the surface mass distribution. 

Under this hypothesis, we have derived analytical expressions for the lensing potential, the deflection angle and the image amplification. Solving the lens equations, we have found that a circularly symmetric Sersic lens produces two images of the source that are merged in the Einstein ring if $r_s = 0$, i.e. the source and the lens centre are perfectly aligned. One of the two images may be deamplified and hence, for some choice of $n$, a Sersic lens does not produce two observable images, an effect that must be taken into account in statistical studies such as lens number counts. No radial critical curve is formed by a circularly symmetric Sersic lens, while the tangential critical curve is the Einstein ring.

Adding a shear term to the lensing potential allows us a first order investigation of the impact of deviations from circular symmetry and to take into account the contribution of nearby lenses. We have solved the lens equations in this case assuming, without loss of generality, that the shear is on\,-\,axis. Due to the breaking of radial symmetry, the number and positions of the images formed depend now both on $r_s$ and $\theta_s$ and not only on $r_s$ as in the symmetric case. We have found that an on axis source gives rise to two images, while two or four images may appear as the source moves away from the lens axis. The presence of the shear also changes the amplification of the images and the shape of the critical curves. The tangential critical curve is deformed into an ellipse, while an inner radial critical curve appears. The size of both curves depends on the Sersic model parameters and decreases with the exponent $n$.

Our results suggest that a Sersic lens can produce up to four images of a distant source if a shear term is added to the potential. This is in agreement with the recent result by Kochanek (2003) who was able to fit constant $M/L$ models to two quadruple QSOs, namely RXJ0911+0551 (\cite{Ba97,Bu98}) and PG1115+080 (\cite{W80,Ietal96}), using the de\,Vaucouleurs profile, that is to say a Sersic lens with $n = 4$. However, a direct comparison is not possible. To fit RXJ0911+0551, one has to use two lensing galaxies modeled with the de Vaucouleurs profile, taking also into account the presence of the cluster which the lenses belong to. A similar situation also holds for PG1115+080 where a single de Vaucouleurs lens is unable to fit the images configuration and the contribution of a nearby cluster is strongly required to reproduce the observable quantities. 

The Sersic profile has been shown to be a better fit to the surface brightness distribution of elliptical lensing galaxies than the de\,Vaucouleurs law. Nonetheless, the de\,Vaucouleurs model is usually used in lensing studies. We have thus investigated the systematic errors induced by this procedure. This analysis has shown that the mass\,-\,to\,-\,light ratio $\Upsilon$ may be seriously under\,- or overestimated (depending on $n$ being lower or larger than 4) when considering constant $M/L$ models. On the other hand, if a dark halo is added to the visible component of the lens galaxy, the use of a de\,Vaucouleurs model instead of the correct Sersic one leads to a wrong estimate of the dark halo projected mass of the dark halo inside the Einstein ring. These considerations could suggest that the use of de\,Vaucouleurs profile to model galaxies that are best fitted by the Sersic law must be avoided. Actually, these results should be treated with great caution since they have been obtained using simulated galaxies whose model parameters have been fixed with a procedure that is by no means unique. In particular, as described in detail in the Appendix, we have used the photometric plane (\cite{KWKM00,G02}) to relate the Sersic parameters $(n, r_e, I_e)$. Actually, this relation has been determined from galaxies belonging to the Virgo and Fornax clusters, while we have extended it to higher redshift galaxies the validity of which is still to be proven. Henceforth, a different prescription to fix the model parameters could be implemented and a detailed study of how this affects the results is needed. However, the procedure we have adopted leads to Sersic parameters describing galaxies which look quite reasonable so that we are confident that our main results are trustworthy. Actually, the best strategy is to resort to a careful study of real lens systems in order to check whether the effects we have reported are indeed true, or a consequence of how we have chosen the model parameters.

The analytical expressions of the lensing potential and of the deflection angle here obtained for a circularly symmetric Sersic lens may be easily generalized to the flattened case following the procedure described, e.g., in Keeton (2001). The resulting potential, deflection angle, and amplification may then be used to try to fit the observed lens systems (both double and quadruple) in order to investigate if constant $M/L$ models are viable solutions or whether a dark matter component has to be invoked to reproduce the observable lensing quantities (images positions and flux ratios). Since the Sersic model parameters $(n, r_e, I_e)$ may be directly measured from detailed photometry, the only unknown lens quantity is the mass\,-\,to\,-\,light ratio $\Upsilon$ and hence we can study eventual correlations among $\Upsilon$ and the photometric parameters. On the other hand, we can assume that lensing galaxies obey the same scaling relations (such as the photometric plane and the fundamental plane) as the low redshift systems. If we model the lens as the sum of a luminous Sersic component and a dark halo, we can reduce the degeneracy among model parameters excluding those solutions which lead to Sersic parameters not allowed by the quoted scaling relations. Moreover, if the time delays among the images were measured, this approach should reduce the systematic error in the determination of the Hubble constant through the time delay method (\cite{Ref64}). On the other hand, statistics of gravitational lensing, such as lenses number counts and distribution of images angular separation, are known to be powerful tools to discriminate among different cosmological models. The modeling of the lenses plays a key role in this kind of analysis and it is thus very important to use realistic lens models. The Sersic lens we have studied may thus be a first step towards a reanalysis of the available data in order to better investigate if it is possible to reconcile them with constant $M/L$ models and if the constraints on the cosmological parameters are changed by the use of more realistic models of elliptical galaxies. These questions will be addressed in a series of forthcoming papers.

\begin{acknowledgements}
We thank an anonymous referee for having suggested us to investigate the systematics discussed in Sect.\,6 and for the comments that have helped to improve the presentation.
\end{acknowledgements}

\appendix

\section{Choosing the lens parameters}

Let us now explain the way we choose the Sersic parameters in order to have a lens system at redshift $z_l$ which is as realistic as possible. To this aim, let us first consider the Milky Way. Its visible component may be modeled as the sum of a bulge with total mass $M_{{\rm bulge}}$ and an exponential disk with scalelenght $R_d$ and local surface density $\Sigma(R_0) \equiv \Sigma_{\odot}$, in which $R_0 = 8.5$\,kpc is the Sun distance to the galactic centre. We set\,:

\begin{displaymath}
M_{{\rm bulge}} = 1.3 \times 10^{10} \ M_{\odot} \ , \ 
\Sigma_{\odot} = 54 \ M_{\odot} \ {\rm pc^{-2}} \ .
\end{displaymath} 
We now build a galaxy with total intrinsic luminosity given as\,:

\begin{equation}
M_{{\rm tot}} =  -2.5 \log{\left ( \frac{M_{{\rm bulge}}} + M_{{\rm disk}}{\Upsilon} \right )}
\label{eq: mtot}
\end{equation}
with $M_{{\rm tot}}$ the galaxy total absolute magnitude and $M_{{\rm disk}} = 2 \pi R_d^2 \ \Sigma_{\odot} \ \exp{(R_0/R_d)}$ the total mass of the Milky Way\footnote{Having chosen the Milky Way as reference for the lens mass scaling is not completely justified since our Galaxy is a spiral one, while we are considering elliptical galaxies as lenses. However, this does not introduce any systematic error in the results since what we really need is a realistic value for the mass of a galaxy. To this aim, we could take as reference any galaxy provided that the mass of its different components are well measured. We have thus chosen the Milky Way as reference only because its mass has been better determined.} disk. The apparent total magnitude of this galaxy at redshift $z_l$ is then\,:

\begin{equation}
\mu_{{\rm tot}}(z_l) = M_{{\rm tot}} + 25 + 5 \log{\left [ \frac{D_L(z_l)}{{\rm Mpc}} \right ]} + K(z_l)
\label{eq: mutotzl}
\end{equation}
with $K(z)$ the $K$\,-\,correction (estimated by interpolating the values reported for elliptical galaxies in \cite{P97}) and $D_L$ the luminosity distance that, for the flat $\Lambda$CDM model, is\,:

\begin{displaymath}
D_L(z) = \frac{c}{H_0} (1 + z) \int_{0}^{z}{\frac{dz'}{\sqrt{\Omega_{\Lambda} + \Omega_m (1 + z')^3}}} \ .
\end{displaymath}
On the other hand, $\mu_{{\rm tot}}$ is related to the three Sersic parameters as follows\,:

\begin{equation}
\mu_{{\rm tot}} = -2.5 \log{\left ( 2 \pi n r_e^2 I_e \int_{0}^{\infty}{x^{2n - 1} {\rm e}^{-b(x - 1)} dx}\right )}
\label{eq: mutot}
\end{equation}
with $r_e$ expressed in $arcsec$. Let us now reduce the Sersic model parameters space using the {\it photometric plane} as found by Khosroshahi et al. (2000; see also \cite{G02}) by fitting the Sersic law to a set of near\,-\,infrared observations of elliptical and S0 galaxies\,:

\begin{equation}
0.172 \ \log{r_e} - 0.069 \ \mu_0 = \log{n} - 1.18 \ ,
\label{eq: phplane}
\end{equation}
in which $r_e$ is in kpc and $\mu_0$ is the central surface brightness related to $I_e$ as follows\,:

\begin{equation}
I_e = {\rm dex}\left [ - \frac{\mu_0 + 1.0857 b(n)}{2.5} \right ]
\label{eq: iemuzero}
\end{equation}
with ${\rm dex}(x) = 10^x$. We have now all we need to describe our procedure of choosing the Sersic parameters for a realistic lens system. 

\begin{enumerate}

\item{Choose a value for the mass\,-\,to\,-\,light ratio $\Upsilon$ and use Eq.(\ref{eq: mtot}) to estimate the absolute magnitude of the galaxy.}

\item{Compute the apparent total magnitude at redshift $z_l$ using Eq.(\ref{eq: mutotzl}).}

\item{For a fixed value of $n$, solve Eqs.(\ref{eq: mutot}), (\ref{eq: phplane}) and (\ref{eq: iemuzero}) with respect to $r_e$ to get the value of the effective radius in kpc and convert it to $arcsec$.}

\item{Estimate $I_e$ as\,:

\begin{equation}
I_e = \frac{10^{-\mu_{tot}(z_l)/2.5}}{2 \pi n r_e^2  \int_{0}^{\infty}{x^{2n - 1} {\rm e}^{-b(x - 1)} dx}} \ .
\label{eq: ieest}
\end{equation}}

\end{enumerate}

This simple procedure ensures us that the lens models we are using to investigate quantitatively the main lensing properties of the Sersic profile are meaningful so that the results we have obtained may be considered reliable. However, it is worth to note that we have implicitly assumed that the photometric plane coefficients in Eq.(\ref{eq: phplane}) do not change with redshift. Actually, this assumption has still to be verified, but we are confident that our main results are not seriously affected by revisions of the photometric plane coefficients. One might also resort to a different procedure, not using Eq.(\ref{eq: phplane}). For instance, one could follow the first two steps of our procedure, then fix both $n$ and $I_e$ and finally evaluate $r_e$ solving Eq.(\ref{eq: mutot}). However, following this prescription, we find that the higher the lens redshift, the larger $r_e$ in complete disagreement with what we observe (and with common sense). This is not the case for our prescription. A detailed and quantitative study of how the results are affected by the procedure adopted to fix the model parameters is nonetheless needed and will be performed in a forthcoming paper.

\end{document}